# Integration of topological insulator Josephson junctions in superconducting qubit circuits

Tobias W. Schmitt[1,2], Malcolm R. Connolly[3,4,*], Michael Schleenvoigt[1], Chenlu Liu[3], Oscar Kennedy[4], José M. Chávez-Garcia[5], Abdur R. Jalil[1], Benjamin Bennemann[1], Stefan Trellenkamp[6], Florian Lentz[6], Elmar Neumann[6], Tobias Lindström[7], Sebastian E. de Graaf[7], Erwin Berenschot[8], Niels Tas[8], Gregor Mussler[1], Karl D. Petersson[5], Detlev Grützmacher[1,2] & Peter Schüffelgen[1,*]

[1] Institute for Semiconductor Nanoelectronics, Peter Grünberg Institute 9, Forschungszentrum Jülich & Jülich-Aachen Research Alliance (JARA), Forschungszentrum Jülich and RWTH Aachen University, Germany.
[2] JARA-Institute for Green IT, Peter Grünberg Institute 10, Forschungszentrum Jülich and RWTH Aachen University, Germany.
[3] Blackett Laboratory, Imperial College, London, United Kingdom.
[4] London Centre for Nanotechnology and Department of Physics and Astronomy, University College London, London, United Kingdom.
[5] Microsoft Quantum Lab Copenhagen and Center for Quantum Devices, Niels Bohr Institute, University of Copenhagen, Copenhagen, Denmark.
[6] Helmholtz Nano Facility, Forschungszentrum Jülich, Jülich, Germany.
[7] National Physical Laboratory, Teddington, United Kingdom.
[8] MESA+ Institute, University of Twente, Enschede, The Netherlands.

*Corresponding Authors:
Peter Schüffelgen, p.schueffelgen@fz-juelich.de
Malcolm R. Connolly, m.connolly@imperial.ac.uk

## Abstract

The integration of semiconductor Josephson junctions (JJs) in superconducting quantum circuits provides a versatile platform for hybrid qubits and offers a powerful way to probe exotic quasiparticle excitations. Recent proposals for using circuit quantum electrodynamics (cQED) to detect topological superconductivity motivate the integration of novel topological materials in such circuits. Here, we report on the realization of superconducting transmon qubits implemented with $(Bi_{0.06}Sb_{0.94})_2Te_3$ topological insulator (TI) JJs using ultra-high vacuum fabrication techniques. Microwave losses on our substrates with monolithically integrated hardmask, used for selective area growth of TI nanostructures, imply microsecond limits to relaxation times and thus their compatibility with strong-coupling cQED. We use the cavity-qubit interaction to show that the Josephson energy of TI-based transmons scales with their JJ dimensions and demonstrate qubit control as well as temporal quantum coherence. Our results pave the way for advanced investigations of topological materials in both novel Josephson and topological qubits.

## Introduction

Topological insulators (TIs) provide an exciting sandpit for exploring novel physics and developing the next generation of electronic devices[1]. One of the most exotic concepts in topological matter are Majorana zero modes (MZMs), quasiparticles with non-Abelian exchange statistics that are predicted to emerge in hybrid devices comprising s-wave superconductors (S) and TIs[2]. Alongside the fundamental interest in detecting MZMs, they could form robust qubits that are immune to most sources of decoherence[3]. $(Bi,Sb)_2Te_3$ TIs have a large helical gap of up to 300 meV[4, 5], making them a particularly attractive platform for realizing MZMs. Over the past decade, various signatures of MZMs have been theoretically proposed in S-TI hybrids[6, 7] and subsequently demonstrated experimentally[8, 9, 10, 11, 12]. Conclusive evidence for MZMs and their non-Abelian statistics, however, will require sophisticated manipulation and measurement protocols[13, 14], so the integration of TIs into advanced quantum circuits is highly desirable.

Circuit quantum electrodynamics (cQED) is the leading architecture for performing state readout and control of superconducting qubits with metal-oxide Josephson junctions (JJs)[15]. By probing microwave transitions and coherent oscillations of JJ-based circuits, cQED is also emerging as a tool for materials research and has been employed to investigate a range of JJs on new systems[16, 17, 18, 19, 20, 21]. Applications of cQED for the detection of topological superconductivity in S-TI-S junctions include spectroscopic signatures[22], and using transmon qubits as parity meters for demonstrating non-Abelian braiding of MZMs in topological qubits[13, 14].

Integrating TIs into hybrid cQED circuits, however, creates additional demands on the microwave compatibility as the environment needs to have low dielectric losses. While $(Bi,Sb)_2Te_3$ TIs can be grown by molecular beam epitaxy on low-loss silicon substrates[23], their sensitivity to ambient conditions imposes further challenges on fabrication. A complete ultra-high vacuum (UHV) fabrication process, based on monolithically integrated dielectric masks for selective area growth (SAG) and stencil metallization, has recently been presented to achieve a scalable, deterministic patterning of TI nanoribbon–superconductor networks[12]. The microwave compatibility and quantum coherence of as-fabricated S-TI hybrid devices, however, has not yet been confirmed. In this work, we experimentally demonstrate quantum coherence in S-TI-S JJs by fabricating and operating a TI-based transmon qubit. We verify the microwave compatibility of our *in situ* fabrication techniques by studying dielectric losses of microwave cavities fabricated on top of our typical substrates. Our results serve as a proof-of-principle for the integration of TIs into superconducting circuits and pave the way for more complex TI-based cQED devices for investigating topological superconductivity and the physics of MZMs.

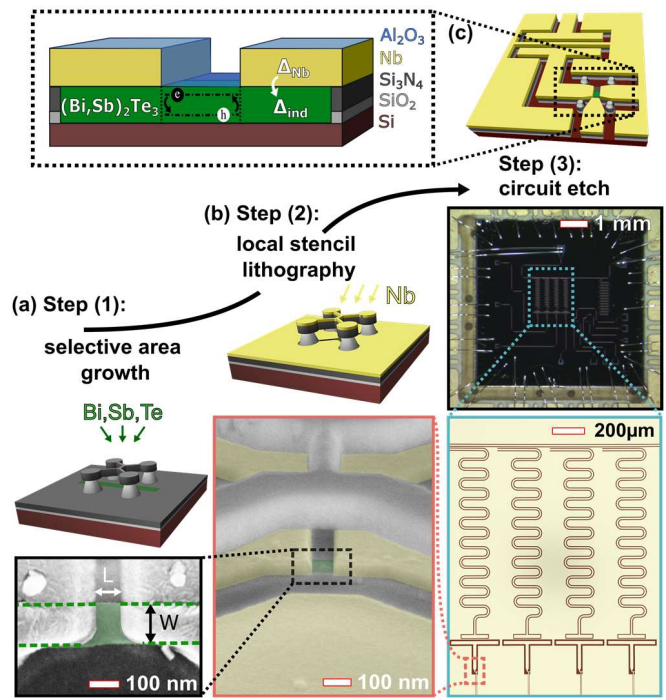

**Figure 1) Integration of topological insulator Josephson junctions in superconducting quantum circuits.** **(a)** Schematic illustrates selective area growth of BST in a trench with sample rotation [step (1)]. Bottom false-colored scanning electron micrograph (SEM) shows a BST nanoribbon (green) of width W. Two Nb electrodes deposited in step (2) are separated by length L. **(b)** Schematic depicts the Nb (yellow) deposition from a fixed angle [step (2)]. False-colored SEM at an angle shows the JJ and the stencil mask used to shadow a part of the BST nanoribbon from the Nb (bottom). **(c)** Top right schematic illustrates the transmon circuitry produced by etching the Nb [step (3)]. Inset (top left) depicts the stack structure in the JJ region, the induced superconductivity in the TI under the Nb electrodes and transport in the JJ via Andreev bound states. Optical micrograph (middle) shows a wire-bonded chip and a false-colored SEM (bottom) the four qubits and readout cavities.

# Results

## Scalable integration of TI Josephson junctions into transmon qubits

Our wafer-scale integration process is illustrated in Fig. 1. We first deposit a selective area growth mask on a silicon wafer and pattern four dry-etched ~2 μm long nanotrenches with controllable widths (here ~100 nm) that define the position and shape of the TI nanoribbons. A second stencil mask is then prepared locally over the nanotrenches to shadow a section of the TI during metal deposition. The substrate preparation is followed by three key steps: In step (1) we grow nanoribbons of $(Bi_{0.06}Sb_{0.94})_2Te_3$ (BST) via molecular beam epitaxy selectively in the nanotrenches with substrate rotation ensuring continuous TI growth under the stencil mask. In step (2) we coat the sample surface with niobium everywhere except for the small area shadowed by the stencil mask. Maintaining UHV conditions ensures clean Nb/TI interfaces with reproduceable properties, previously hampered by *ex situ* processing[24, 25, 26]. Finally, a stoichiometric $Al_2O_3$ passivation layer is deposited to protect the exposed TI during subsequent processing. In step (3) the microwave circuitry is etched in the Nb film and the stencil mask is removed. Fig. 1c shows the final four-qubit chip with the cQED readout circuitry. Each qubit comprises a T-shaped island shunted to ground via a TI nanoribbon. The nonlinear inductance of the TI JJ generates an anharmonic qubit spectrum whose lowest states $|0\rangle$ and $|1\rangle$ form the qubit basis. Similar hybrid transmon qubits have been realized with JJs made from III-V semiconductor nanowires[27, 28], 2DEGs[29], graphene[30, 31, 32] and $MoTe_2$[33]. In the transmon limit $E_J \gg E_C$, the qubit frequency is given by $E_{01} = hf_{01} \approx \sqrt{8 E_J E_C}$, where $E_C$ is the charging energy of the qubit island and $E_J \approx \hbar I_c/2e$ is the Josephson energy of the JJ with $I_c$ being the critical current. We design devices such that $E_C/h \approx 240$ MHz and $E_J$ is varied by fabricating nanoribbons with different lengths L and widths W. Quarter wavelength (λ/4) co-planar waveguide cavities with frequencies in the range of $f_r$ = 6.9 - 7.3 GHz are capacitively coupled to the qubits for readout and manipulation. Using electrostatic simulations of the qubit-cavity capacitance, we estimate the qubit-cavity coupling strength of our devices to be $g_{cav}/2\pi \approx 90$ MHz assuming a qubit frequency of $f_{01}$ = 6 GHz[34]. This is in good agreement with our experimental results from two-tone spectroscopy measurements yielding $g_{cav}/2\pi \approx 93$ MHz (see Supplementary section B). Due to the qubit-cavity interaction at low photon numbers the cavities are shifted from their bare resonance frequency. This dispersive shift is given by $\tilde{f}_r - f_r = \chi = g_{cav}^2/\Delta$, where $\Delta$ is the detuning between the qubit and the cavity. In the first measurement run, we observe such shifts for six qubits on two different chips and extract their Josephson energies $E_J$ by assuming a fixed $g_{cav}/2\pi$ = 93 ± 9.3 MHz. Fig. 2a plots $E_J$ as a function of the JJ dimensions, related via the expression $We^{L/\xi}$, where $\xi$ = 110 nm is the superconducting coherence length of *in situ* fabricated Nb-BST thin film JJs extracted in previous work[12]. Using the DC transport data from that work (blue data, Fig. 2a), we were able to tune $E_J$ over almost two orders of magnitude and realize qubits in the transmon regime by adjusting the geometries of the Josephson junctions in the range of L ≈ 90 – 150 nm and W ≈ 110 – 140 nm. Such control of $E_J$ on a design parameter is an important step for demonstrating the viability of TIs as scalable qubit platform.

## Two-tone spectroscopy

Fig. 2b shows the two-tone spectroscopy performed on a qubit on TI-mon chip 1, obtained by varying the power $p_d$ and frequency $f_d$ of the qubit drive and reading out the cavity at a fixed readout tone $f_c$. The drive tone excites the qubit into a mixed state, causing a qubit state-dependent push on the shifted cavity frequency $\tilde{f}_r$. This shift is detected by measuring the demodulated response $V_H$ of the readout tone. At low power in Fig. 2b (green trace), we observe the $|0\rangle \rightarrow |1\rangle$ transition and establish a qubit frequency of $f_{01}$ = 6.30 GHz. At higher drive power we observe a 2-photon process $f_{02}/2 \approx 6.15$ GHz driving the $|0\rangle \rightarrow |2\rangle$ transition. The difference in frequency of these two transitions allows us to estimate the anharmonicity of the qubit $\alpha/h = 2 (f_{02}/2 - f_{01}) \approx 300$ MHz. In contrast to highly-transmitting channels in hybrid transmons[17], which typically show $\alpha < E_C$, our device is closer to the metal-oxide limit ($\alpha \approx E_C$). This behavior might be caused by diffusive bulk contributions which are superimposed on the transparent ballistic surface modes in TIs[12] and the absence of a helical surface mode in confined TI nanowires[4].

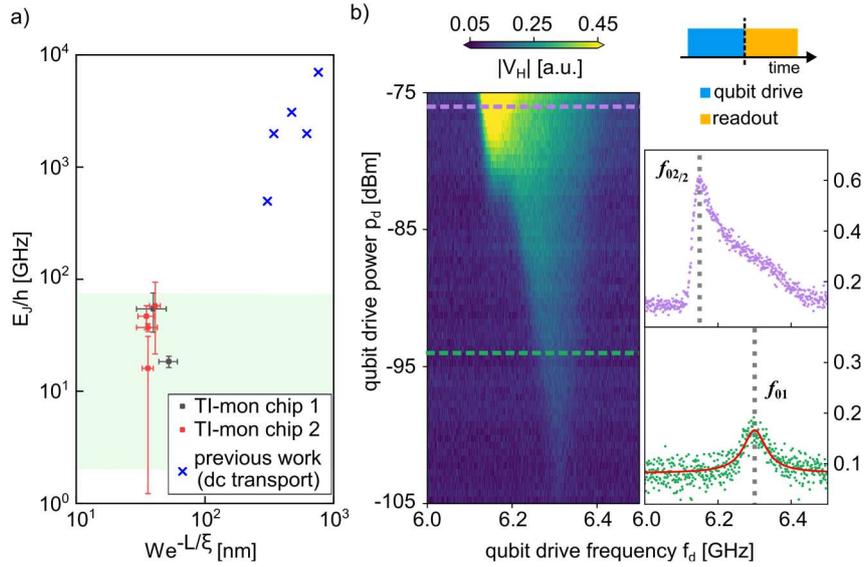

**Figure 2) Characterization of TI transmon qubits. a)** Josephson energies of six qubits from two different chips as a function of their Josephson junction geometries, all in an accessible regime for transmons (qubit frequency at ~2-12 GHz, indicated in green). The energies have been extracted from the power dependent push on the respective resonator during a first measurement run on these devices. Previously published data obtained from DC transport on similarly fabricated Nb-BST JJs are shown as comparison (blue)[12]. **b)** Heterodyne readout voltage measured as a function of qubit drive frequency $f_d$ and power $p_d$ using a single-sidebanded qubit drive (JJ geometry: W ≈ 120 nm, L ≈ 90 nm). Single traces are taken at -76 dBm (violet) and -94 dBm (green). The frequency $f_{01}$ is determined by a Lorentzian fit to the data at -94 dBm (red solid line) and $f_{02}/2$ is identified with the maximum of the curve at -76 dBm.

## Coherence and control of a TI transmon

Next, we show coherent qubit control of the TI transmon. Using the pulse sequence in Fig. 3a we drive Rabi oscillations of the qubit between the qubit states $|0\rangle$ and $|1\rangle$ by varying the duration $\tau_{Rabi}$ of the excitation pulse. These Rabi oscillations depend on the drive frequency $f_d$ and have their longest period at the qubit frequency $f_d = f_{01}$ (Fig. 3a). At a fixed drive frequency, the period of the oscillations is adjusted by varying drive power $p_d$ (Fig. 3b). These Rabi oscillations have been used to calibrate $\pi/2$ and $\pi$ pulses which are required to investigate the qubit coherence. To probe the phase coherence we prepare the qubit in a coherent superposition and perform a Ramsey interference measurement (Fig. 3c). As shown in the pulse sequence, the qubit is prepared with a $\pi/2$ pulse and allowed to evolve for a time $\tau_{Ramsey}$ prior to a second $\pi/2$ pulse and a readout pulse. In this measurement we observe fringes at finite detuning of the drive frequency from the qubit frequency $\Delta f = f_d - f_{01}$, a signature of coherent precession of the qubit. As the frequency of these oscillations is given by $\Delta f$ a corresponding linear dependence can be observed in the discrete time Fourier transform of the Ramsey data. Additional frequency components might come from the 2-photon transition around $f_{02}/2$ or from excitation of higher levels due to short $\pi/2$ square pulses (~2.3 ns). The dephasing time can be estimated to $T_2^* = 10$ ns by a fit to the Ramsey fringes in Fig. 3c. Next, we measure the energy-relaxation time $T_1$ by first sending a $\pi$ pulse (which excites the qubit from the $|0\rangle$ to the $|1\rangle$ state) and varying the waiting time $\tau_{T1}$ before reading out the qubit state (Fig. 4a). By fitting an exponential decay to the data we extract $T_1 = 28$ ns. The fact that $T_2^* < 2T_1$ indicates that the dephasing is currently not limited by energy relaxation but rather pure dephasing which might, for example, be caused by low-frequency noise[35]. Although these timescales are short compared to mature metal-oxide transmon qubits, they are comparable with recently presented hybrid qubits on graphene[32] and provide a proof-of-principle that topological insulators can be integrated into cQED systems.

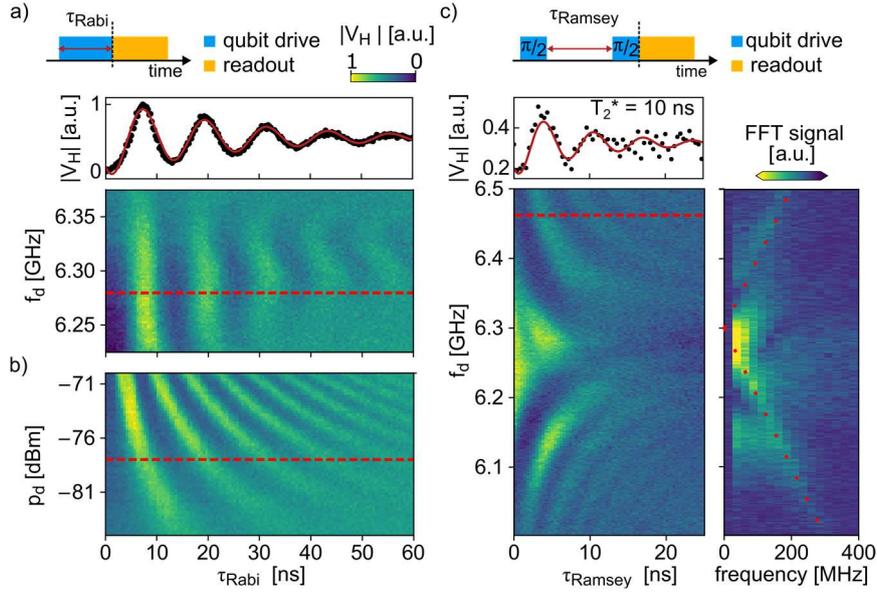

**Figure 3) Coherent control of a TI transmon qubit. a)** Rabi oscillations of a TI transmon qubit measured with the pulse sequence shown at the top. The single line plot shows an average of 10 Rabi measurements taken at a drive frequency $f_d$ = 6.28 GHz and power $p_d$ = -78 dBm with the solid red line being a fitted damped sinusoid. The color plot shows the dependence of the Rabi oscillations on the drive frequency $f_d$. **b)** Power dependence of the Rabi oscillations at a drive frequency $f_d$ = 6.28 GHz. **c)** Coherent precession of the qubit in a Ramsey measurement at $p_d$ = -75 dBm. The oscillations at finite detuning from $f_{01}$ are shown in the line cut taken at $f_d$ = 6.462 GHz and fitted with a damped sinusoid (solid red line) to extract $T_2^*$ = 10 ns. Discrete time Fourier transform (right) of the Ramsey measurement is taken for each $f_d$ (using the fast Fourier transform algorithm) with dotted red line as a guide to the eye. For better visibility, the mean value of each Rabi and Ramsey measurement is subtracted in the color plots.

**Analysis of microwave losses in SAG substrates**

Parasitic two-level systems (TLS) suspected in amorphous materials or at interfaces have been identified as an ubiquitous loss mechanism in cQED[36]. While our local approach of the stencil lithography mitigates possible dielectric losses in the stencil mask, the superconducting circuit of our TI transmon devices is patterned on top of the global selective area growth mask. To characterize microwave losses from this mask, we fabricated Nb lumped element resonators with a 5 nm $Al_2O_3$ capping layer directly on a typical mask layer (20 nm $Si_3N_4$/5 nm $SiO_2$) (see inset Fig. 4b, Supplementary Fig. S6a). As shown in Fig. 4b, we investigated dielectric losses introduced by these layers in microwave transmission measurements. The resonators show an average single-photon internal quality factor of $(6.6 \pm 0.3) \times 10^4$ (determined using a robust and traceable fit routine[37]) implying that the SAG mask substrates would allow for potential energy relaxation times on the order of $T_1 = Q/2\pi f \approx 2$ µs for a transmon limited by substrate TLS losses. Fitting the power dependent internal quality factor of the test resonators at mK to widely used models of losses from TLS[38] allows us to extract the loss tangent of the resonators due to TLS (see Supplementary section F). We find the average value of $F\tan(\delta_{TLS}^0) \approx 1.1 \times 10^{-5}$ which is only one order of magnitude worse than mature technologies of Al resonators on high resistivity silicon[39].

**Discussion**

Our results imply that TI transmon qubits are currently not limited by TLS losses in the SAG substrate which is promising for future development based on the SAG technique. Imperfect selective growth on our first-generation qubit devices resulted in TI crystals grown on the SAG mask (see white dots Fig.1a) which may add to losses but can be avoided by adjusting the growth temperature[12]. Also, the JJ itself might contribute to short coherence times. As already proposed for InAs nanowire transmon devices[27],

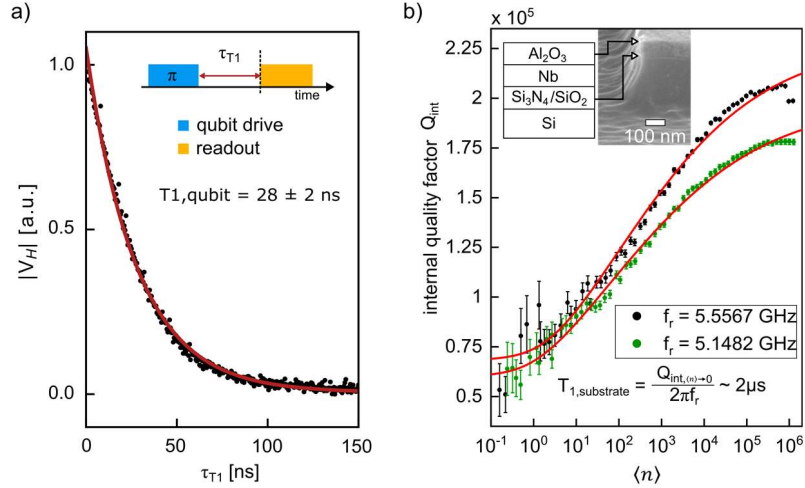

**Figure 4) Energy-relaxation of qubit and substrate losses. a)** Energy-relaxation time measurement of the qubit using the pulse sequence in the inset. Data are averaged over a total of 150 single $T_1$ measurement runs with small temporal variance of ± 2 ns (see Supplementary section E). By fitting an exponential decay to the data (red) we extract $T_1$ = 28 ns. **b)** Internal quality factor of two lumped element resonators as a function of average photon number <n> in the resonator. This is fit to a model of TLS induced losses. Inset is a SEM micrograph showing a cross sectional view of the device (including trenching of the Si substrate) which has been fabricated equivalently to the TI transmon samples

a transparent superconductor-semiconductor interface and a hard superconducting gap is suspected to play an important role for such hybrid qubits. While a high transparency has already been reported on *in situ* fabricated Nb-BST JJs[12], a thorough analysis of the induced superconducting gap in TI nanoribbons might provide a route to improve qubit coherence in future work. Alternative superconductors which are currently investigated for an epitaxial, highly transparent interface to tetradymide TIs[40, 41] may readily replace the Nb used in this work. Moreover, tunability of the qubit frequency by gate[29, 32] or flux[27] control will help to study the sensitivity of qubit coherence to different types of noise and provide a better understanding of losses in these junctions. From a technological point of view, establishing gate-tunable topological JJs[42, 43] in cQED systems promises a suitable approach for voltage-controlled Josephson qubits. While III-V semiconductor nanowire[27, 28] and two dimensional electron gas (2DEG)-based[29] gatemons as well as tuneable couplers[44] demonstrate the viability of this approach, our TI hybrid qubits offer an alternative platform that unites pristine superconductor-semiconductor interfaces, a low-loss host substrate, and deterministic patterning.

In conclusion, we have demonstrated the compatibility of ultra-high vacuum fabricated topological insulator Josephson junctions with cQED. We showed that our device fabrication based on selective area epitaxy and local stencil lithography allows for designing Josephson energies reliably on a cQED compatible substrate and demonstrate qubit control and coherence. This proof-of-concept work confirms the presented technique as a variable and scalable approach to integrate TIs into superconducting circuits which could be used to study topological Josephson junctions and explore novel TI based superconducting qubits, such as gatemons, Majorana transmons[22] or top transmons[13]. Furthermore, it paves the way for read out of topological qubits in TI - superconductor networks[12].

## Methods

### Device fabrication

For the fabrication of the selective area growth mask, a 4'' Si(111) wafer (> 2000 Ωcm) was thermally oxidized in an oxidation furnace for 5 nm of $SiO_2$ prior to deposition of 12 nm of stoichiometric $Si_3N_4$ via LPCVD. In this $Si_3N_4$ layer, trenches were prepared for selective area growth using electron beam lithography (EBL) and reactive ion etching (RIE). The stencil mask was fabricated by deposition of 300 nm $SiO_2$ and 100 nm stoichiometric $Si_3N_4$ via LPCVD. Subsequently, the wafer was diced and we used a negative EBL process and RIE to define the local stencil masks in the top $Si_3N_4$ layer aligned to the nanotrenches in the SAG mask. Prior to growth, the samples were cleaned in piranha ($H_2SO_4 + H_2O_2$) and the $SiO_2$ was isotropically etched using hydrofluoric acid (1%) which released the local stencil mask and passivated the dangling bonds of the Si(111) substrate in the SAG trench. This hydrogen passivation was removed by heating the substrate in an MBE chamber to $T_{sub}$ = 700°C prior to growth. Following the recipe of Kellner et al.[45], a thin film of ~ 20 nm $(Bi_{0.06}Sb_{0.94})_2Te_3$ was selectively grown into the nanotrenches at $T_{sub}$ = 355°C. During this deposition the sample has been rotated (10 rpm). The MBE cells in this chamber are mounted at a polar angle of 32.5° allowing for a continuous TI growth under the stencil mask. For metal deposition, the samples were transferred to another MBE chamber using a portable vacuum chamber keeping the minimum pressure below $10^{-8}$ mbar. Using an electron beam evaporator, a film of 50 nm Nb has been deposited at a rate of 1Å/s. For this deposition, the sample was aligned such that the nanoribbons were perpendicular to the impinging flux. As the stencil mask was wider than designed (due to the sensitivity of a negative EBL process), narrower junctions were realized by continuously moving the sample between ± 7° around this position during the Nb deposition, narrowing the junction length effectively by ~ 60 nm to its planned geometry. The capping layer of 5 nm $Al_2O_3$ has been deposited under sample rotation (10 rpm) using an electron beam evaporator and a stoichiometric $Al_2O_3$ target. Both TI-mon samples have been fabricated in individual, consecutive MBE runs. For the circuit etching a positive EBL and RIE process have been used. For TI-mon chip 1, metallic shorts (owing to resist problems during the microwave circuit etching) have been repaired via focused ion beam (FIB) cutting. For the resonator test sample, the SAG mask deposition (here: 5 nm $SiO_2$/20 nm $Si_3N_4$), the Nb and $Al_2O_3$ deposition as well as the circuit definition has been perform equivalently to the TI-mon samples. Except for the LPCVD depositions and TI MBE steps, all fabrication processes were performed at the Helmholtz Nano Facility (HNF)[46].

**Experimental set-up**

All measurements were performed in cryogen-free dilution refrigerators with a base temperature of T < 50 mK. Details of the attenuation, isolation and amplification are shown in Supplementary Fig. S1.

# Acknowledgments

We gratefully acknowledge helpful discussions with J. Burnett, M. Kjaergaard and Th. Schäpers. We thank A. Hertel and M. Eichinger for their assistance with the time-domain measurements. T. W. Schmitt thanks C. M. Marcus for supporting a stay at the Center for Quantum Devices. This work was supported financially from the German Federal Ministry of Education and Research (BMBF) via the Quantum Futur project "MajoranaChips" (Grant No. 13N15264) within the funding program Photonic Research Germany, the Priority Programme SPP1666, European Union's Horizon 2020 research and innovation programme (grant agreement No 766714/HiTIMe), the department for Business Energy and Industrial Strategy (BEIS) as well as by Germany's Excellence Strategy - Cluster of Excellence "Matter and Light for Quantum Computing" (ML4Q) EXC 2004/1 – 390534769. We moreover acknowledge support by the IVF project "Scalable Solid State Quantum Computing" and by EPSRC EP/L020963/1. T. W. Schmitt acknowledges support from the German Academic Exchange Service (DAAD). K. D. Petersson and J. M. Chávez-Garcia acknowledge support from Microsoft and the Danish National Research Foundation.

# Author contributions

# Supplementary information

## A. Spectroscopy measurement setups

We used two different cryogen-free dilution refrigerator configurations shown in Fig. S1 to collect data for the spectroscopy of TI-mon samples. Both setups were equipped with two coaxial lines for the measurements. The signal lines (red) are heavily attenuated and the read-out lines (blue) are shielded by isolators/circulators which forward the signal while suppressing noise to the sample. The signals were amplified at cryogenic temperatures and also further at room temperature. The initial measurements of low-power cavity pushes shown in Fig. S2a and used for main text Fig. 2a were performed for TI-mon chip 1 in the setup shown in Fig. S1a without the superconducting travelling wave parametric amplifier (TWPA) and the directional coupler. For TI-mon chip 2 the low-power cavity pushes have been measured in the setup in Fig. S1b. We loaded TI-mon chip 1 for a second time in setup S1a with the TWPA and took time-domain data for one of the qubits. While the two-tone spectroscopy shown in Fig. 2d, the Stark shift measurement in Fig. S3, the energy relaxation measurements in Fig. 4a and Fig. S5 as well as the Chevron Rabi measurement in Fig. S4 were taken without pumping the TWPA (= no amplification), all data shown in Fig. 3 were taken with a TWPA pump (= amplification). For the resonator test measurements shown in Fig. 4b and Fig. S6 we used a setup similar to Fig. S1a without TWPA and directional coupler.

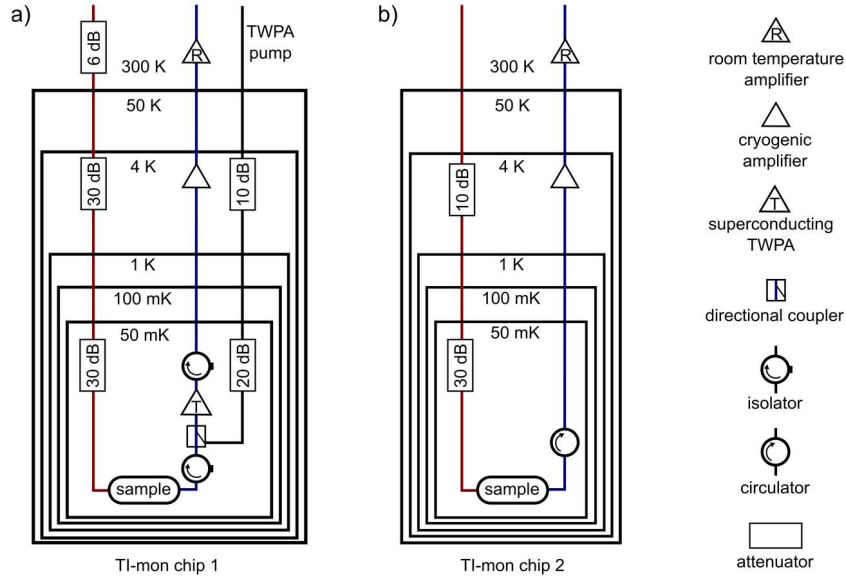

**Figure S1)** Cryogenic setups used for measurements with different attenuation, isolation and amplification on the signal line (red) and read-out line (blue).

## B. Comparison of same qubit for different loads

In this section we briefly discuss the two different measurement runs of a cavity/qubit on TI-mon chip 1. The data of the first load shown in Fig. S2a are used for main text Fig. 2a. We used the low-power dispersive shift of the resonator and the qubit frequency extracted from two-tone spectroscopy data to estimate the qubit-cavity coupling to $g_{cav}/2\pi \approx 93$ MHz. In order to demonstrate qubit control and coherence, we loaded the sample a second time (~20 month after the first load). Compared to the first load, the bare frequency of the cavity decreases by ~ 3.6 MHz which could be an aging effect of the cavity due to oxidation. We measure an increased dispersive shift (from 7.3 to 11.7 MHz) and observe accordingly an increased qubit frequency in two-tone spectroscopy (from 5.98 GHz to 6.3 GHz). This corresponds to a change in the Josephson energy $E_J/h$ from 18.6 GHz to 20.7 GHz (or ~ 4 nA in the critical current of the junction) which seems reasonable for a semiconductor based Josephson junctions.

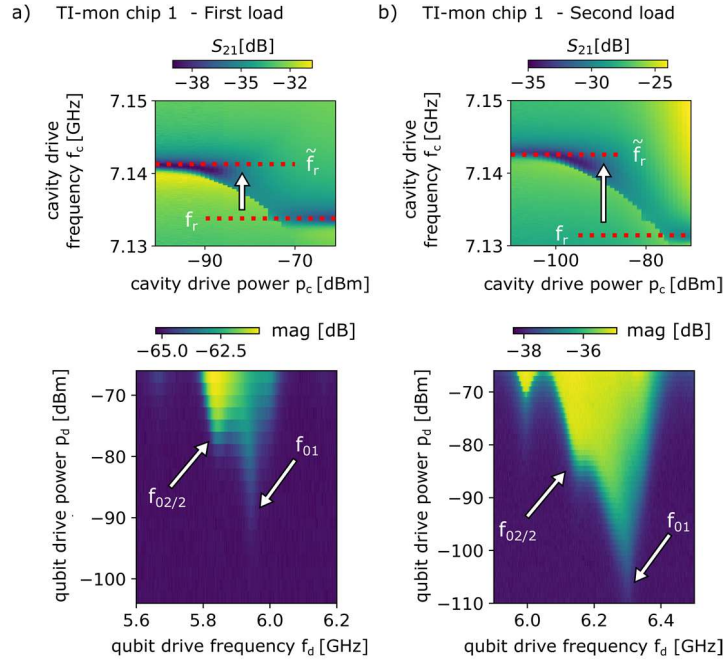

**Figure S2) Comparison of first (a) and second (b) load of TI-mon chip 1.** The power dependence of the cavity (top) shows the dispersive shift in VNA transmission measurements for both experimental runs. Using an external microwave source and a VNA, we also measure the power dependent two-tone spectroscopy of the qubit (bottom). In the second load, both transition (namely $f_{01}$ and a two-photon peak $f_{02/2}$) are shifted up by ~300 MHz.

## C. Stark shift

Since the qubit is coupled to the cavity its occupation with photons acts on the dressed qubit states. This results in a photon number dependent push on the qubit known as ac Stark shift[47]. It is measured in two-tone spectroscopy with overlapping, continuous drive/readout tones by keeping the qubit excitation power fixed at $p_q$ = -101 dBm but varying the readout power $p_r$ (and hence the number of photons in the cavity). The corresponding measurement showing the ac Stark shift of the $f_{01}$ transition can be seen in Fig. S3. In order to get a good signal-to-noise ratio we used a readout power of $p_r$ = -89 dBm for time-domain measurement. As these measurements are done in pulsed, non-overlapping spectroscopy the high readout pulse power does not cause a significant Stark shift.

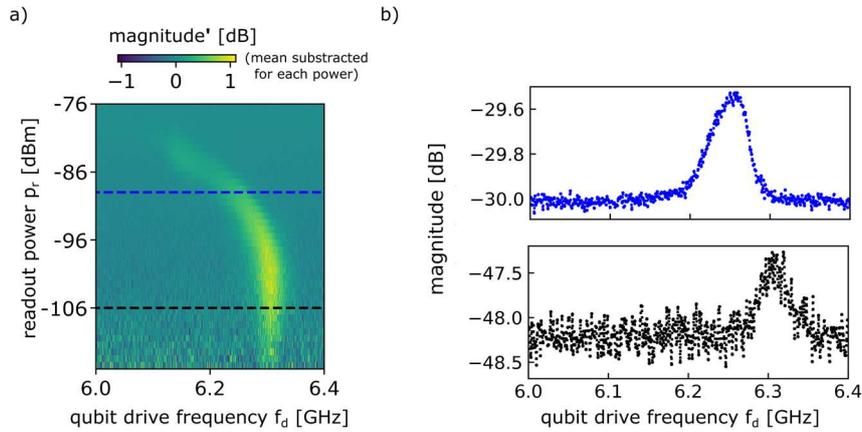

**Figure S3) Stark shift of the qubit $f_{01}$ transition. a)** Using an external microwave source and a VNA, we measure the readout power dependent two-tone spectroscopy at a qubit drive power of $p_q$ = -101 dBm. The mean value at each readout power has been subtracted for visibility. **b)** Two single two-tone spectroscopy measurements at readout power $p_r$= -89 dBm (blue) and $p_r$= -106 dBm (black) without mean value subtraction.

## D. Power dependence of Rabi measurements

As shown in Fig. S4, we also measure the frequency dependence of the Rabi oscillation (similar to main text Fig 3a) for different drive pulse powers. The pulse sequence for these measurement is the one shown in Fig. 3a. Additional features present at higher drive powers disappear at lower drive powers where we can recognize the qubit frequency $f_{01} = 6.3$ GHz by the maximal Rabi period.

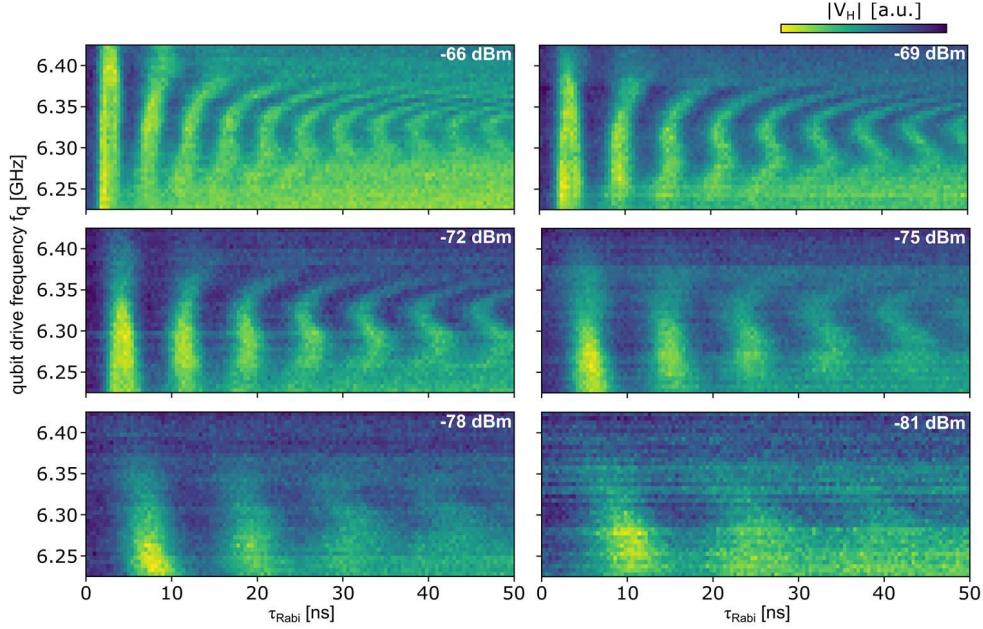

**Figure S4) Frequency dependence of Rabi measurements for different drive powers.** As the time resolution for these measurements was chosen finer than the AWG sampling rate we binned and averaged data taken with the same physical waveforms.

## E. Temporal dependence of T1

The energy-relaxation times in transmon qubits have been observed to vary over time[48]. Therefore, we took a total of 150 consecutive $T_1$ measurements over a period 378 minutes using a drive frequency of $f_q = 6.28$ GHz and a drive power of $p_q = -78$ dBm (Fig. S5a). These data are taken with the same pulse sequence shown in main text Fig. 4a. We find an average of $T_1 = 28 \pm 2$ ns with a small temporal variance (Fig. S5b,c). This low temporal variance allows to average over all 150 $T_1$ measurement yielding main text Fig. 4a. For future devices that include gate or flux tunability it would be interesting to study spectral and temporal dependence of $T_1$ to investigate possible energy-relaxation channels[48].

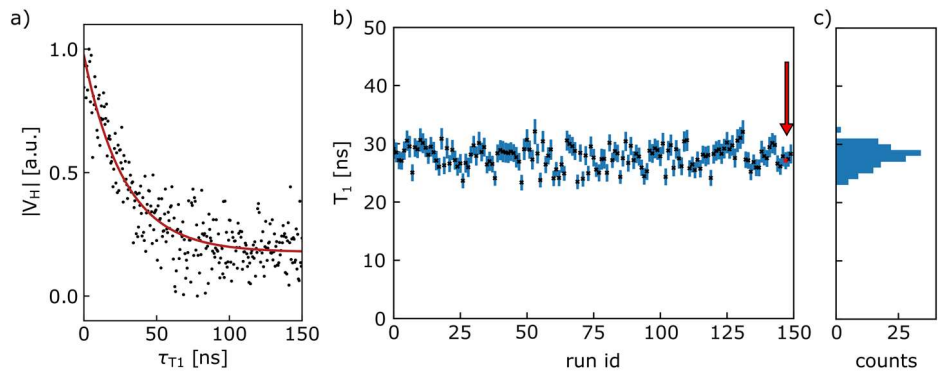

**Figure S5) Temporal dependence of $T_1$. a)** A single energy-relaxation measurement with fitted exponential decay to extract $T_1$. **b)** Extracted $T_1$ times from a total of 150 energy-relaxation measurements. The measurement shown in (a) is indicated in red. **c)** Histogram of all $T_1$ times with 1 ns bins.

## F. Power dependence of the TLS loss

Fig. S6a shows a lumped element test resonator which we use for studying TLS losses. A typical transmission measurement at one power and the corresponding fit to extract the Q-factor are shown in Fig. S6b. In order to determine the TLS losses of our resonators we analyze the power-dependent internal quality factor data in the framework of the standard tunneling model (STM)[38, 39]. At strong microwave fields $\langle n \rangle \gg n_c$, where $n_c$ is the critical photon number for the saturation of TLS, it is described by the following power law:

$$ 1/Q_i = \frac{F \tan(\delta_{TLS}^0)}{(1 + \langle n \rangle / n_c)^\alpha} + 1/Q_{i,0} \tag{S1}$$

In this model, power-independent losses are covered by $Q_{i,0}^{-1}$ and the exponent α describes the strength of the power saturation of the Q-factor. An α < 0.5 is an indication of TLS interactions (with α = 0.5 being the result of the non-interacting STM). For our resonators, we find the power-independent contribution to the total loss rate $Q_{i,0}^{-1} \approx 4.7 \times 10^{-6}$ and α ≈ 0.24. The TLS loss tangent $F\tan(\delta_{TLS}^0)$ is found from the fit to eq. S1.

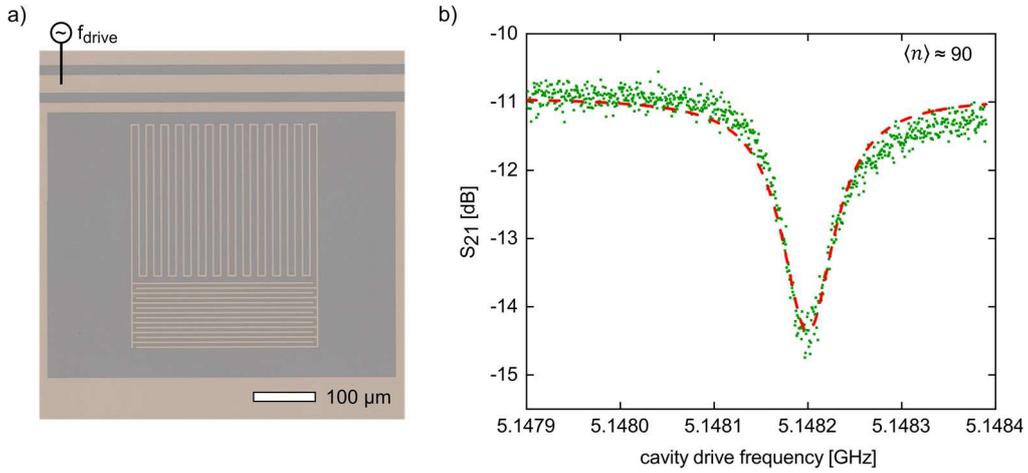

**Figure S6) Power dependent measurements on test resonators. a)** Microscope image of a lumped element resonator. **b)** Measurement of the microwave transmission magnitude ($S_{21}$) of a lumped element resonator taken at ⟨n⟩ ≈ 90 using a VNA. A fit (red) is used to extract the internal quality factors.